\documentstyle[multicol,psfig,epsf,aps]{revtex}
\begin{document}
\draft
\title{\bf{Steric repulsion and van der Waals attraction between 
flux lines in disordered high $T_c$ superconductors}}
\author{Sutapa Mukherji and Thomas Nattermann} 
\address{Institut f\"{u}r Theoretische Physik,
 Universit\"{a}t zu K\"{o}ln, D-50937 K\"{o}ln, Germany}
\maketitle
\begin{abstract}
We show that in anisotropic or  layered superconductors
impurities induce a van der Waals attraction between flux lines.
This attraction together with the disorder induced repulsion 
may change  the low $B$ - low $T$ phase diagram
significantly from that of the pure thermal case considered 
recently by Blatter and Geshkenbein [Phys. Rev. Lett. {\bf 77}, 4958 (1996)].

\end{abstract}
\pacs{74.60.Ge}
\widetext
\begin{multicols}{2}

Conventional type--II superconductors show in addition to
the flux repulsing Meissner state a second superconducting (Abrikosov)
phase in which the magnetic induction ${\bf B}$ enters the material in the
form of quantized flux lines (FLs) 
which form a triangular lattice.
Each FL carries a unit flux quantum $\Phi_0= hc/2e$. The Abrikosov 
lattice is characterized by a non--zero shear modulus $c_{66}$, which 
vanishes at the upper
and lower critical field, $H_{c_2}$ and $H_{c_1}$, where  continuous 
transitions to the normal and the Meissner state, respectively, occur.
In his mean--field solution Abrikosov treats FLs as stiff rods. 
Close to the lower critical field 
$H_{c_1}$, their interaction becomes exponentially weak
and hence the FL density $l^{-2}=B/\Phi_0$ vanishes as 
$|\ln\tilde{h}|^{-2}$ where $\tilde{h}=(H-H_{c_1})/H_{c_1}$ denotes the 
reduced field strength \cite{tink}.

Thermal fluctuations roughen  the FLs resulting in a possible
melting of the Abrikosov lattice 
close to $H_{c_1}$ and $H_{c_2}$, respectively, because of the 
softening of $c_{66}$.
This applies in particular to high--$T_c$ materials with their
elevated transition temperatures and their pronounced layer
structures \cite{blatrev}. 
At present, it is not clear whether the transition
to the normal phase at high field happens in these materials 
via one or two transitions.
However, melting of the FL lattice has clearly been observed 
experimentally \cite{gammel}. 

At low fields a first order melting transition to a
liquid phase and a change in the critical behavior of $B$ 
has been predicted some time ago by Nelson\cite{nel}. 
 Quantitatively the influence of thermal fluctuations
is described by a thermal length scale 
$L_T=\Phi_0^2/(16\pi^2T)\approx 2{\rm cm}K/T$ \cite{fish}.
$L_T$ has a simple physical meaning: an isolated flux line of length 
$L_T$ shows a thermal mean square displacement of the order of the London 
penetration length $\lambda$. 
Besides a shift of $H_{c_1}$, large scale thermal fluctuations lead 
close to $H_{c_1}$ to an 
entropic repulsion 
$\sim \left( \lambda^2/L_Tl\right)^2$ between FLs which dominates over the 
bare interaction for small $\tilde{h}$ and hence $B \sim \tilde{h}$ 
\cite{nel}. 
(Here we measure all FL 
interactions in units of $\varepsilon_0 = (\Phi_0/4\pi\lambda )^2=
L_{T} T/{\lambda}^2$).

More recently, Blatter and Geshkenbein \cite{blatt} found that in 
anisotropic or layered
superconductors short scale fluctuations give rise also 
to an {\it attractive} 
van der Waals (vdW) interaction \cite{brandt}. 
For FLs separated by a 
 distance $R$ the strength of this interaction is of the order of
$\lambda ^6 /L_T(d+\varepsilon R)R^4 $. $\varepsilon^2 =m/M \ll 1$
denotes the anisotropy of the material with $m$ and $M$ the effective masses
parallel and perpendicular to the $ab$ plane, and $d$ the interlayer spacing.
$\lambda$ and $\lambda /\varepsilon$ are then the screening
lengths parallel and perpendicular to the layers, respectively.

The competition among  the bare, the entropic and the vdW interactions 
leads to an interesting phase diagram at low $B$ values.
In particular, Blatter and Geshkenbein \cite{blatt}
 find at low $T$ a first order transition
between the Meissner and the Abrikosov phase.

So far fluctuation effects have been discussed for a clean superconductor.
It is well known, however, that in type--II superconductors FLs have to be
pinned in order to 
prevent dissipation from their motion under the influence
of an external current. 
Therefore, besides the thermal fluctuations 
one has to take into account the effect of disorder. 
Randomly distributed pinning centers lead indeed
to a destruction of the Abrikosov lattice \cite{larkin},
 but as has been recently shown,
for not too strong disorder FLs may form a (Bragg--) glass phase
which is characterized by quasi long--range order of the FL
lattice \cite{bragg}. 
For low $B$ this phase undergoes a melting transition
to a pinned liquid state.
Inside this phase, disorder induced effects are
expected to dominate over those of thermal fluctuations for
sufficiently low $T$.
It is therefore the aim of the present paper to consider the influence 
of the disorder fluctuation induced forces between the FLs,
and to study the low $B$ phase diagram. In particular we 
find, that the latter
deviates substantially from that 
found in Ref. \cite{blatt}  for pure systems. The rest of the paper is
organized as follows. We first reconsider the disorder mediated 
steric repulsion between the FLs and then derive the disorder induced 
vdW interaction. Finally we discuss the phase diagram at low $B$
and $T$.

Let us start with the {\it steric repulsion}, 
which results from the long wave length
fluctuations of FLs. Using a simple scaling 
argument, it was found in Ref. \cite{natter}, 
that the disorder dominated steric repulsion between FLs
is of the order
\begin{eqnarray}
{\lambda^{2/\zeta}}{L_{\rm dis}}^{-2}l^{2-2/\zeta}= 
( T_{\rm dis}/\epsilon_0 \xi)^2
(\xi /l)^{2/{\zeta }-2}\label{eq1}
\end{eqnarray}
where the roughness exponent $\zeta $ of a single FL in a random potential
is about 5/8 in $D=3$ dimensions \cite{forrest}. Here $\xi$ is the 
correlation length in the superconducting planes
and $L_{\rm dis}$ denotes
a disorder--related length scale 
with a similar meaning as $L_T$.
At low temperatures,
$ L_{\rm dis}\approx L_{T}
T\kappa^{1/\zeta-2}/ T_{\rm dis}$, where
$\kappa=\lambda/\xi$, denotes the Ginzburg-Landau parameter. $T_{\rm dis}$
is defined in (\ref{dist}), and it is identical with ${\tilde
  T}_{\rm dp}^{\rm s,iso}$ of  ref. \cite{blatrev}.
The steric interaction results from the fact, that the FL is confined to a 
cylindrical cage, formed by  its six nearest neighbors. It therefore cannot 
gain the full energy decrease a free FL can obtain from the disorder.

This argument has been later put into question \cite{fieg}, 
since the 
FL might overcome the averaged repulsive potential $(\lambda /l)^{D-1}$ of 
the cage and hence leave it to follow its optimal path. Naively, 
this should be the case if   
$(\lambda /l)^{D-1} < {\lambda^{2/\zeta}}{L_{\rm dis}^{-2}}l^{2-2/\zeta}$
or  $D>D_c=(2/\zeta (D))-1$. Such an argument is indeed correct for thermal 
fluctuations, where $\zeta = 1/2$ in all dimensions: for $D>3$ 
the steric repulsion becomes ineffective since the entropy gain from 
leaving the cage outweighs the repulsion \cite{seung}. 

In the case of disorder, the situation is however different. At low $T$ a FL 
which leaves its cage can only increase its energy gain if it follows the 
optimal path. Let us 
neglect for the moment the repulsive interaction between the lines. Since the 
optimal path for a line with a given initial point has a river delta like 
shape, 
two lines separated initially by a distance $nl$ will  merge
after a transient region of length $ (nl)^{1/\zeta} \ll L$.
Let us now switch on again the repulsion. Since 
the lines follow the same path, repulsion leads to an energy increase of the
order $L$. This has to be compared with the free energy gain by 
following the optimal path instead of the path inside the cage, which is of 
the order $L^{2\zeta -1}$ and hence smaller than the repulsion
since $\zeta<1$.
Thus, a line leaving 
its cage would increase its energy instead of decreasing it unlike 
the thermal case, where lines can avoid each
other and still gain entropy. Consequently, the simple argument of 
Ref. \cite{natter}
 applies which results in (\ref{eq1}).

The above argument can indeed be derived in 
 a somewhat more formal way by using a renormalization group (RG)
 analysis of interacting FLs \cite{mukh}.
In the large wavelength limit, the FLs of length $L$ 
are described by the Hamiltonian
\noindent\begin{eqnarray}
{\cal H}=\!\int^{L}_0\! dz\sum\limits_{i=1}^N\{ \frac{\varepsilon_l}{2}
{\dot{\bf R}}_i^2+ 
\varepsilon_{\rm pin}({\bf R}_i,z)+
 \sum\limits_{j\not= i}v_0\delta_{\lambda}
({\bf R}_{ij})\}\label{hamil},
\end{eqnarray}
where ${\dot{\bf R}}_i=\partial {\bf R}_i(z)/\partial z$ and
${\bf R}_{ij}={\bf R}_i-{\bf R}_j$.
$({\bf R}_i(z),z)$ denotes the position vector of the i--th
FL, $\varepsilon_l$ its stiffness constant which is of the
order $\varepsilon_0$ in the long wavelength limit. 
The random pinning potential 
fulfills ${\overline{\varepsilon_{\rm pin}({\bf R},z)
}}=0$ and \cite{blatrev}
\begin{eqnarray}
{\overline{\varepsilon_{\rm pin}({\bf R},z)\varepsilon_{\rm pin}({\bf 0},0)
}} =(T^3_{\rm dis}/\varepsilon_0\xi^2)\ \delta (z)\tilde{k}({\bf R}/\xi
)\label{dist} 
\end{eqnarray}
with $\tilde{k}({\bf x})=1$ for $x\ll 1$ and 
$\tilde{k}({\bf x})\approx (1/x^2)\ln x$
for $x\gg 1$, respectively.  $T^3_{\rm dis}$
is proportional to the impurity concentration. 
$\delta_{\lambda}({\bf R})$
represents a $\delta$--function, smeared out over a scale $\lambda$
and $v_0\approx 4\pi\lambda^2\varepsilon_0$. Since we are interested
in the diluted limit $l\gg\lambda$, the RG flow of
$\Delta$ and $v$ can be obtained by 
 considering only two interacting lines as in Ref.\cite{mukh}. 
Close to $D=2$ dimensions, we find 
$(\partial v/\partial l)=2(1/\zeta -1)v$ and hence 
$v$ remains  a relevant perturbation even above
two dimensions, in contrast to the results obtained in Ref. \cite {fieg}.
 We, therefore, conclude that the result (1)  is valid for all $D$. 
It is worth remarking here, that in contrast to the present case,
 a {\it single} FL in a cylindrical
cage does indeed show a depinning transition from this potential in $D>2$ 
dimension \cite{hwa}.

Next we consider the attractive {\it van der Waals interaction}.
We start with the case of {\it extreme anisotropy} $\varepsilon =0$, in
which the Josephson coupling between the layers can be
 neglected. The repulsive
interaction between two pancake vortices 1 and 2 separated by a distance
$R$ in the same layer is then larger than their attractive interaction
in different layers by a factor $\lambda /d\ (\gg 1)$.
 In considering the interaction of 
different FLs consisting of pancake vortices we therefore restrict 
ourselves to pancakes in the same layer. Since their interaction is
$2d\ln (\xi /R)$, the interaction of two pancake dipoles 
resulting from the displacements ${\bf u}_1,\ {\bf u}_2$ of pancakes
1 and 2 in the same layer is $U_{12}=-(2d/R^2)[2({\bf u}_1\cdot
\hat{n})({\bf u}_2\cdot
\hat{n})-{\bf u}_1\cdot {\bf u}_2]$, where $\hat n$ is the unit vector in the 
direction connecting the two FLs.
A displacement ${\bf u}_1$ will indeed induce a displacement ${\bf u}_2$
because of the force ${\bf f}_{12}=-\partial U_{12}/\partial {\bf u}_2$.
The actual response of the pancake 2 is limited
by the elastic force ${\bf f}_{22}\sim -(d
{\bf u}_2/\lambda^2)\,
\ln (\lambda \pi /d)$ resulting from the other pancakes of FL 2. 
Here we have used the stiffness constant in the decoupled limit 
\cite{blatrev}, $\varepsilon_l(k)=
(\varepsilon_0/2\lambda^2k^2)\,\ln (1+\lambda^2k^2)$ for $k\approx\pi /d$.
With ${\bf f}_{12}+{\bf f}_{22}=0$ and averaging over different
configurations of ${\bf u}_1$, the vdW--interaction per unit
length is
\begin{equation}
V_{\rm vdW}\sim-\frac{\lambda^2}{R^4}
\frac{1}{\ln\pi\lambda /d}{\overline{\left< 
{\bf u}^2\right>}} \label{vdw}
\end{equation}
Inserting the result for the short wavelength 
thermal fluctuations 
$\left< {{\bf u}^2}\right>=(2T/\varepsilon_0d)\,
[\lambda^2/\ln (\pi\lambda /d)]$, we obtain the vdW attraction given 
in  Ref. \cite{blatt}, apart from a numerical factor.

In the case of an impure superconductor ${\overline{\left< {{\bf
          u}^2}\right>}}$ 
is given by the mean square displacement of a pancake in a potential which is 
a superposition of a 
parabolic elastic part $\varepsilon_0d({\bf
  u}/2\pi\lambda)^2\ln{(\lambda\pi /d)}$, resulting from the
dispersive elastic constant for $k\approx \pi/d$,
and a random potential $\int_0^d dz\varepsilon_{\rm pin}({\bf
  u},z)$.
Rewriting  (\ref{hamil}) for $L\approx d$ in dimensionless
quantities ${\bf r}={\bf R}/\xi$ and $t=z/d$, it is easy to see, that
the ground state displacement ${\bf u}_0/\xi$ is only a functional of
$(\Delta(\pi\lambda/d))^{1/2}{\tilde \varepsilon}_{\rm pin}({\bf r},t)$. 
${\tilde \varepsilon}_{\rm pin}({\bf r},t)$ is the dimensionless random
potential of mean zero and variance unity and 
\begin{equation}
\Delta({\rm x})=\Delta_0 \frac{{\rm x}}{\ln^2(1+{\rm x}^2)},
\ \Delta_0=
\left(\frac{T_{\rm
      dis}\kappa^2}{\varepsilon_0\lambda}\right)^3 \label{expo}
\end{equation} 
Thus at $T\lesssim T_{\rm dis}$, 
${\overline {{\langle \bf u}^2\rangle}}\approx{\overline {{\bf u}_0^2}}=
\xi^2 f(\Delta(\pi\lambda/d))$.
Perturbation theory applies for $\Delta\ll 1$ and gives 
$f(\Delta)\approx \Delta^{\eta}$
with $\eta =1$. For larger $\Delta$ 
no exact result is known. Imry--Ma 
arguments and a variational treatment  give
 $\eta =1/2$ with a logarithmic correction \cite{engel}, whereas  
we find
$\eta\approx 4/9$  from simulations.
For BSCCO with $\varepsilon=1/300$, $\lambda\approx 2000 {\AA}$, 
$\xi \approx 20{\AA},\  d=15 {\AA},\ {\rm and} 
\  T_{\rm dis}=45K$ 
\cite{blatrev},  $\Delta_0\approx 91$ and 
$\Delta\approx 261$, i.e. we are in the non-perturbative regime.
 In Ref. \cite{ertas} a much larger value of $T_{\rm dis}=210K$ was
 reported which results $\Delta_0\approx 9.261\times 10^3$ and 
$\Delta\approx 2.6\times 10^4$.  The final form for the disorder induced vdW
interaction in the decoupled limit is therefore
\begin{eqnarray}
V_{\rm vdW}^{(\rm dis)}\approx & & -\frac{\kappa^{-2}\varepsilon_0}{
\ln \pi\lambda /d}\left(
\frac{\lambda}{R}\right)^4 \left[ 
\Delta_0
\left(\frac{\pi\lambda/d}{\ln^2(\pi
    \lambda/d)}\right)\right]^{\eta}.\label{eqnvd}
\end{eqnarray}
From (\ref{vdw})- (\ref{eqnvd}), we conclude that,  
in the decoupled limit,  the vdW interaction
will be dominated by disorder fluctuations for $T\lesssim {\tilde
  T}_{\rm dis}$. For $\Delta\gg 1$ (and $\eta=1/2$), ${\tilde T}_{\rm
  dis}=\kappa (T_{\rm dis}^3 d/\varepsilon_0 \lambda^2)^{1/2}$,
whereas for $\Delta\ll 1$, ${\tilde T}_{\rm dis}=\kappa^4 (T_{\rm
  dis}/\varepsilon_0 \lambda)^3 \varepsilon_0 \lambda/\ln[\pi
\lambda/d]$. With the above values this results in ${\tilde T}_{\rm
  dis}=8K$ for $T_{\rm dis}=45K$ \cite{blatrev}, and ${\tilde T}_{\rm
  dis}=83K$ for  $T_{\rm dis}=210K$ \cite{ertas}.

Next we consider the {\it continuous anisotropic} case $\varepsilon >0$.
Inspection of the perturbation theory \cite{blatt}
 shows, that if the disorder induced interaction between FLs is
 neglected the vdW interaction
can be written in the following form 
\begin{eqnarray}
V_{\rm vdW}=-\frac{1}{4T}\left(\frac{\Phi_0^2}{4\pi}\right)^2 
\int_0^{\pi/d} \frac{dk}{2\pi}[V_{\rm xx}^{\rm int}(k,R)]^2 k^4 \nonumber \\
\times (C_T^2(k)+
2C_T(k)C_{\rm dis}(k)),\label{eqnew1}
\end{eqnarray}
where the interaction between two vortex segments is \cite{blatt}
\[
V_{\rm xx}^{\rm int}(k,R)=-\frac{1}{2\pi R}\frac{\varepsilon}{\lambda
{\sqrt{1+\lambda^2 k^2}}}K_1(\varepsilon R{\sqrt{1+\lambda^2
  k^2}}/\lambda),\]
 with $K_\nu$ the $\nu$th order modified Bessel function. 
The two different correlations $C_T$ and $C_{\rm dis}$ are defined as
$
C_T(k)=\overline{
\langle{\bf{u}}_k{\bf{u}}_{-k}\rangle-\langle{\bf{u}}_k\rangle
\langle{\bf{u}}_{-k}\rangle}=T/(\varepsilon_l(k)k^2)$ \cite{schulz} 
and
$C_{\rm dis}(k)=
{\overline{\langle {\bf u}_k\rangle\langle {\bf u}_{-k}\rangle}}$.
The first 
contribution in  (\ref{eqnew1}) resulting from $C_T(k)$,  yields
the thermal vdW interaction which  was 
considered in Ref. \cite{blatt}. The second contribution comes 
from the disorder and vanishes as the impurity concentration vanishes. To
find $C_{\rm dis}(k)$ we have to distinguish the cases $\lambda k\ll 1$
and $\lambda k\gg 1$. For $\lambda k\ll1$ we may neglect the
dispersion of $\varepsilon_l(k)$ and use the known result for
${\overline {{\bf u}_0^2}}$ on large scales \cite{blatrev}. 
 This gives $C_{\rm dis}(k)\sim
{\xi^2}{k^{-1}}(\Delta(\lambda k))^{2 \zeta/3}$. On the contrary, for
$\lambda k\gg 1$, the strong dispersion of $\varepsilon_l(k)$
essentially decouples pancake vortices in different layers. Hence
$C_{\rm dis}\approx {\overline {{\bf u}_0^2}}d$, apart from logarithmic
corrections. Using the asymptotic behavior of the Bessel function
$K_1(z\rightarrow 0)\sim z^{-1}$  
 and $K_1(z\rightarrow\infty )\sim e^{-z}$,
 it is easy to see, that the main contribution to
the second integral in (\ref{eqnew1}) comes from large $k$, $k\lesssim
{\pi}/{(d+\varepsilon R)}$. In a convenient interpolation form the
disorder induced van der Waals interaction for
$\lambda<R<\lambda/\varepsilon$ is then given by  
\begin{equation}
V_{\rm vdW}^{(\rm dis)}(R)\approx - \frac{\varepsilon_0}{\kappa^{2}}
(\Delta_0\frac{\lambda}{d})^{\eta} 
(\frac{\lambda}{R})^4 \frac{d}{d+\varepsilon R} 
(\ln \frac{\pi \lambda}{d+\varepsilon
  R})^{-1-2\eta}\label{ddep}
\end{equation}
Eqn. (\ref{ddep}) is the main result of this paper. For
$\varepsilon R\ll d$ it changes over to (\ref{eqnvd}). 
Thus similar to the thermal case, the vdW attraction decays as 
${R^{-4}}$ or ${R^{-5}}$ for $\varepsilon R<d$ or
$\varepsilon R>d$, respectively. 
Eqn. (\ref{ddep}) can indeed be written in the form of the thermal vdW
interaction \cite{blatt} if we replace there $T$ by ${\tilde T}_{\rm
  dis}$.

In the last part we  analyze the
phase diagram at low $T$ and ${H}\agt H_{\rm c1}$.
The bare repulsion between the flux lines $2 K_0(R/\lambda)$  and
their vdW attraction result in a  
minimum of their interaction energy 
at a distance 
$R_{\rm min}\approx \alpha \lambda$ $(\ll l)$. For not too low
temperatures $\alpha $ is about $20$  and only weakly $T$
dependent. The same applies to the width of the minimum which is of
the order of $
\beta \lambda$ with $\beta\approx 10$. Since the vdW  attraction is
strongly distance dependent, its main contribution comes from those
configurations where the line pair is at a distance $R_{\rm
  min}$. To lowest order in $\Delta_0$, we can estimate the {\it average}
vdW interaction by considering the configurations of a single line in
the absence of any FL  interaction. 
With $u\!\approx \! \lambda (L/L_{{\rm T},{\rm dis}})^{\zeta}$ 
 for the displacement
of a single FL, we find from $u\approx l$ for the mean distance
$L_\parallel$ between  two line segments  reaching a minimum
$L_\parallel \approx L_{\rm T,dis}(l/\lambda)^{1/\zeta}$. 
The length $L_s$ of
the segment over which the line stays in the minimum follows from the
same argument as $L_s\approx L_{\rm T,dis}\ \beta^{1/\zeta}$. Thus, 
the contribution
of the vdW attraction to the Gibbs free energy density is of the order
\begin{equation}
\frac{1}{l^2} V_{\rm vdW}(R_{\rm min})\frac{L_s}{L_\parallel}\approx
V_{\rm vdW}(R_{\rm
  min})(\frac{\lambda}{l})^{2+1/\zeta}\frac{\beta^{1/\zeta}}{\lambda^2},
\end{equation}
which is much larger than the vdW interaction at the mean distance $l$.
For  thermal fluctuations $(\zeta=1/2)$ the mean vdW attraction has
therefore the same $l$ dependence as the entropic repulsion. In this
case, the  result can also be obtained by mapping the problem onto
$2d$- Bosons \cite{bg}.
 For disorder induced fluctuations, however, the vdW
attraction decays faster than the steric repulsion.
The total Gibbs free energy density 
for $\varepsilon R_{\rm min}\gg d$ can be written 
in the following form
\begin{eqnarray}
G(x;H,T,T_{\rm dis})\approx\frac{\varepsilon_0}{\lambda^2x^2}\{
zK_0(x) 
+&&\frac{(\gamma_T-\delta_T)}{x^2}+\frac{\gamma_{\rm dis}}{x^{6/5}}-
\nonumber\\
- &&\frac{\delta_{\rm dis}}{x^{8/5}}-\tilde{h}\ln{\kappa}
\}\label{gibbs}
\end{eqnarray}
which has to be minimized with respect to $x=l/\lambda$. Here $z=6$ is
the number of nearest neighbors for a triangular lattice. 
$\gamma_T\approx
9.08(T/\varepsilon_0\lambda )^2$ \cite{blatt}
 and $\gamma_{\rm dis}\approx
c_{\rm dis}(T_{\rm dis}/\varepsilon_0\lambda
)^2\kappa^{4/5}$ 
denote the strength of the entropic
and disorder dominated steric repulsion respectively. 
Expression (\ref{gibbs}) has to be considered as an interpolation
between the regimes dominated by the bare interaction at high $B$
and the different fluctuation induced interactions at low $B$, 
respectively.
The prefactors of the terms following from the vdW interactions are 
$\delta_T\approx c_{\rm T}
\frac{\beta^2}{\alpha^5}(1/\varepsilon )
(T/(\varepsilon_0\lambda \ln^2\frac{\pi}{ \varepsilon
  \alpha}))$ and $\delta_{\rm dis}\approx  
{\tilde c}_{\rm dis} \frac{\beta^{8/5}}{\alpha^{5}}(1/\varepsilon)
(T_{\rm dis}/\varepsilon_0\lambda)^{3\eta}\kappa^{-2+6\eta} 
{|\ln \frac{\pi} {\alpha\varepsilon} |}^{-(1+2\eta)}
(\lambda/d)^{-1+\eta}$. 

The precise determination of the coefficients $c_{\rm T}$, $c_{\rm
  dis}$, ${\tilde c}_{\rm dis}$ as well as of $\alpha$ and $\beta$ is
beyond the scope of this paper. To find them would require an RG
treatment similar to that performed in \cite{seung} for short range
repulsive FLs in the thermal case. In this case, instead of singling
out a typical distance $R_{\rm min}$, the contributions of the vdW
interaction from all distances will be taken into account according to
their statistical weight. We postpone this study to a future
publication \cite{av} and discuss here the phase diagram (Fig. 1) 
only qualitatively.

\hbox{\psfig{file=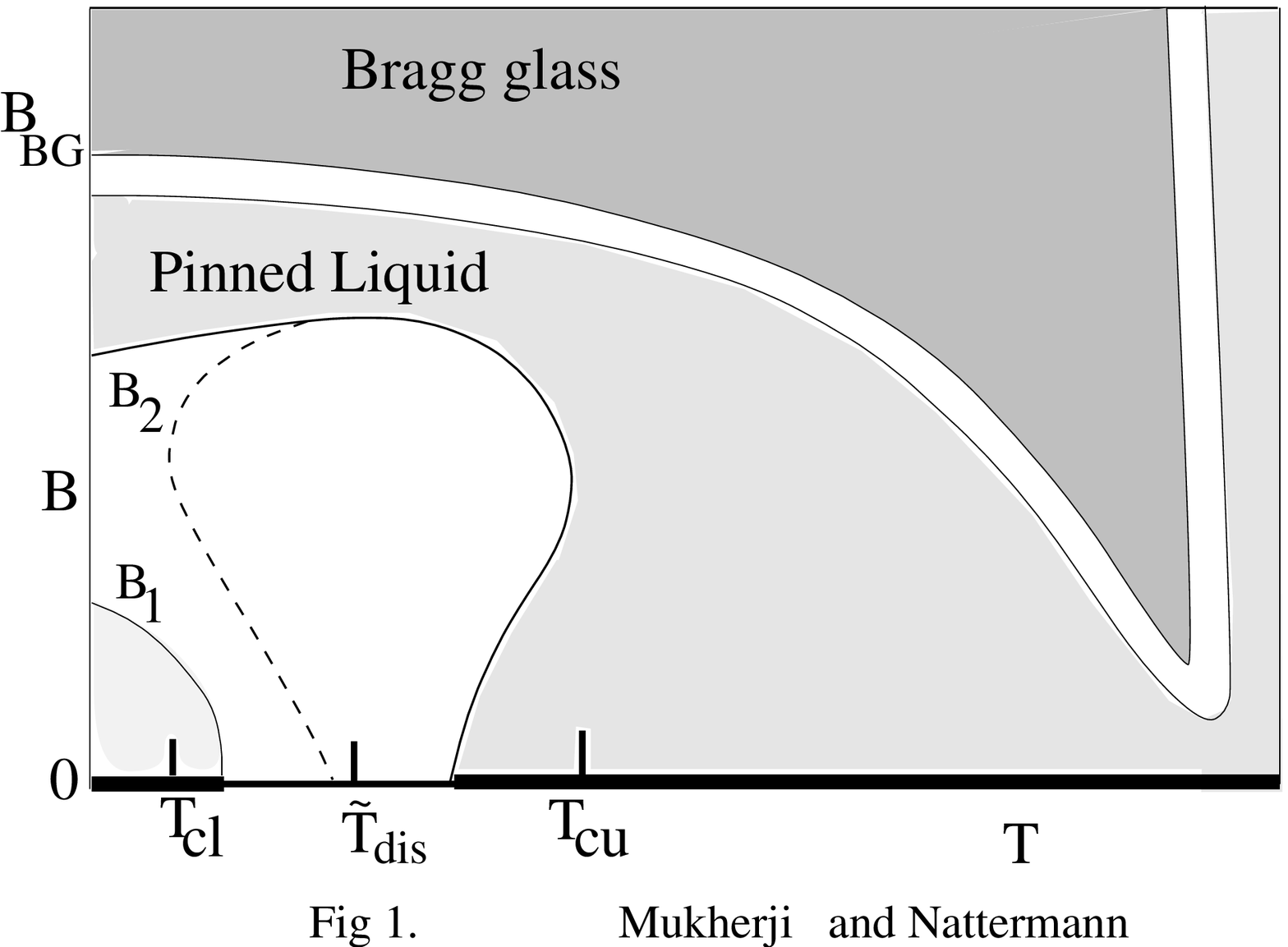,width=2.4in,angle=0}}
\begin{figure}
\narrowtext
\caption{The low- $B$ phase diagram of an impure anisotropic
  superconductor. Thick lines correspond to continuous
  transitions. White regions indicate the coexistence regimes of the
  adjacent phases. $B=0$ corresponds to the Meissner phase. For
  sufficiently weak disorder and low $T$ a low density pinned gas phase is
  separated from a high density pinned liquid phase by a first order
  transition where $B$ jumps from $B_1$ to $B_2$. 
  With increasing disorder, the $B_1$ and $B_2$ lines
  merge as indicated by the dashed line, leading to a critical point
  at $T_{\rm cl}$. For large disorder $T_{\rm cl}$ moves towards
  $T_{\rm cu}$ and the first order transition disappears completely.}
\end{figure}
For $T=0$, to begin with, we find from (\ref{gibbs}) that for
increasing $H$ the transition from the Meissner phase to the pinned liquid
phase is continuous 
 since the steric repulsion dominates over the vdW interaction for  large
$x=l/\lambda$ and hence $B \sim {\tilde h}^{5/3}$. Increasing $H$ further,
for weak enough disorder (but $\Delta>1$) the vdW attraction will
dominate over the steric repulsion in an intermediate range of $x$
resulting in a discontinuous transition from a dilute to a dense
pinned liquid phase. 
At even higher $B$ a second discontinuous transition to the
Bragg-glass phase takes place. At finite but low $T$ this picture is
essentially unchanged and will then cross over smoothly to the thermal phase
diagram studied in \cite{blatt}. For stronger disorder the steric
repulsion dominates at $T=0$ for all values of $B$ and the
discontinuous transition disappears. The first
order transition between the  low (or zero) $B$ phase 
and the dense pinned liquid phase is
now shifted to higher $T$ and will disappear for sufficiently strong
disorder completely. The latter is most likely the situation in BSSCO with
impurity concentration corresponding to $T_{\rm dis}=45K$ or higher.

The coexistence regime is well below the transition to the Bragg glass
phase where the quasi long range order of the lattice persists. This
transition line is beyond the reach of this analysis and 
corresponds to a much higher value of $B_{\rm BG}$ \cite{bragg}.

In conclusion, we have obtained a disorder induced van der Waals
attraction and a steric repulsion 
 between the flux lines in anisotropic or layered superconductors. 
A qualitative analysis of the low field phase diagram for such 
impure system  predicts a rich phase diagram with a first order
transition between two pinned gas/liquid phases for not too strong
disorder. For larger disorder the first order transition disappears
and $B \sim (H-H_{\rm c1})^{5/3}$ decreases continuously as
$H\rightarrow H_{\rm c1}$.

We thank G. Blatter and V. Geshkenbein for sending us their paper prior
to publication and E. H. Brandt for critical reading of the manuscript. 
We are grateful to V. Geshkenbein for pointing out a mistake in the 
original version of the paper as well as several very useful suggestions. 
We acknowledge the financial support from Sonderforschungsbereich SFB
341. T. N. also acknowledges support from German Israeli Foundation.

\end{multicols}

\end{document}